\documentclass[letterpaper, 10 pt, conference]{ieeeconf}
\IEEEoverridecommandlockouts
\usepackage{amsmath}
\usepackage{pgfplots}
\pgfplotsset{compat=1.18}
\usepackage{url}

\usepackage{mathtools}
\usepackage{amssymb}
\usepackage{cite}

\newtheorem{asm}{Assumption}
\newtheorem{lem}{Lemma}
\newtheorem{thm}{Theorem}
\newtheorem{crl}{Corollary}
\newtheorem{rem}{Remark}

\newcommand{\N}{\ensuremath{\mathbb{N}}}
\newcommand{\R}{\ensuremath{\mathbb{R}}}
\newcommand{\norm}[1]{\left\lVert#1\right\rVert}

\title{Closed-loop Properties of Value Iteration in Continuous Time with Approximation Errors}
\author{Francisco Moreno-Mora and Stefan Streif
\thanks{F. Moreno-Mora and S. Streif are with the Professorship of Automatic Control and System Dynamics, Chemnitz University of Technology, 09107 Chemnitz, Germany (e-mail: \protect\url{francisco.moreno@etit.tu-chemnitz.de}; \protect\url{stefan.streif@etit.tu-chemnitz.de}).}}

\begin{document}

\maketitle
\thispagestyle{plain}
\pagestyle{plain}

\begin{abstract}
Value iteration is an important tool for approximating solutions of infinite-horizon optimal control problems, but in practice it is never carried out exactly: value functions are represented by finite-dimensional approximators, and the iteration is stopped after finitely many steps. This paper analyzes which closed-loop guarantees remain under these limitations for continuous-time nonlinear systems. We consider a value iteration scheme in which each iteration solves a finite-horizon optimal control problem with the previous iterate as terminal cost, and the associated controller applies the optimal input of this problem over each horizon in a sampled-data fashion. We show that if the convergence tolerance and the approximation error are small enough, the controller asymptotically stabilizes the system on sublevel sets of the computed iterate, and provide a suboptimality bound. This condition consists of pointwise inequalities that can be checked for the computed iterate, without establishing convergence of the iteration, and is a first step towards sample-based certification. We further establish inherent robustness of the controller with respect to model errors and recover classical results for the linear-quadratic regulator. Numerical studies on a linear-quadratic problem and on a nonlinear system with known optimal value function are consistent with the guarantees and show that the suboptimality bound is conservative.
\end{abstract}

\section{Introduction}

Infinite-horizon optimal control problems characterize the best achievable performance through the optimal value function, which, under suitable assumptions, also serves as a Lyapunov function for the optimally controlled system. Solving the associated Hamilton--Jacobi--Bellman equation is, however, intractable for general nonlinear systems. Approximate dynamic programming (ADP) circumvents this by iteratively updating approximations of the value function or of the policy \cite{lewis2009reviewADP,lewis2013RLandADPforFeedback}. Furthermore, in the context of model predictive control (MPC), stability for finite prediction horizons is commonly ensured by terminal costs satisfying a Lyapunov decrease condition \cite{mayne2000constrainedMPCstab_optim,chen1998LQR_terminal}. Such terminal costs are hard to construct for nonlinear systems, and approximate value functions are a natural candidate. In discrete time, combining value iteration (VI) with MPC has led to stability and suboptimality guarantees that account for approximation errors and permit shorter prediction horizons \cite{beckenbach2022approximateIHpredictive,moreno2023MPCwithApproxVI}. The present analysis provides a continuous-time foundation for such schemes.

Among its two main schemes, policy iteration generally requires an initially stabilizing policy, whereas VI does not, which often constitutes a superfluous requirement in view of the fact that the goal of optimal control is usually to stabilize the system. Accordingly, convergence and stability of VI, including the effect of approximation errors, have been studied extensively \cite{al2008ADPconvergence,heydari2016aproxVI,heydari2018aproxVI}.

Value iteration is seldom carried out exactly. Value functions on continuous state spaces are represented by finite-dimensional approximators, every iteration involves numerical integration and optimization, and the procedure is stopped after finitely many steps. Each of these steps introduces an error, and the controller that is eventually deployed is derived from an imperfect iterate. The central question is therefore not only whether the iterates approach the optimal value function, but what an imperfect iterate guarantees for the closed loop: is the associated controller stabilizing, how robust is it, and how far is its cost from the optimal one?

Most of this theory is formulated in discrete time even though physical systems have continuous-time dynamics, and discretizing a continuous-time system to apply it may not yield the expected results. As the sampling interval shrinks, the cost accumulated over a single transition vanishes, whereas approximation and numerical errors do not. Error conditions stated relative to this cost thus become increasingly restrictive \cite{tallev2019robustQlearningTimeDiscr}. Moreover, guarantees for the discretized problem need not carry over to the underlying continuous-time system \cite{nesic1999sampledDTapprox,Vrabie2007ContiADP}. Continuous-time ADP has been developed both for policy iteration \cite{Vrabie2007ContiADP,beckenbach2024stabilizingRLsampled} and for VI \cite{xiao2024convergenceVIcont,zhang2018VI_cont_H_infty}. In particular, \cite{xiao2024convergenceVIcont} proves convergence of continuous-time VI under a contraction assumption and shows that, with bounded approximation errors, the iterates reach a neighborhood of the optimal value function. Convergence of the iterates and stability of the feedback derived from them are, however, distinct properties: closed-loop guarantees require the approximation and convergence errors to be related to a Lyapunov decrease. For discrete time, such relations have been established in \cite{heydari2016aproxVI,heydari2018aproxVI,lincoln2006relaxingDP}. To the best of our knowledge, no result relates the approximation error and the convergence tolerance of continuous-time VI to the stability, robustness, and suboptimality of the associated controller.

This work addresses this gap. We consider a VI scheme in which each iteration solves a finite-horizon optimal control problem with the previous iterate as terminal cost. The associated controller for the final iterate is  then used in a sampled-data manner with the already mentioned horizon as sampling period \cite{fontes2001sampledMPC}. We show that if the convergence tolerance, measured relative to the cost accumulated over the horizon, and the approximation error, measured relative to the minimal cost over one horizon, fulfill a bound, this controller asymptotically stabilizes the system, also between sampling instants, on sublevel sets of the iterate from which its optimal trajectories end in the domain. This condition involves only the last iteration step and can be checked directly, without first establishing convergence of the iteration. In general, however, it can only be evaluated at sampled states. We therefore see the present results as a first step towards sample-based certificates for approximate value functions: they identify which quantities must be certified, and relative to which costs. The same condition yields an explicit decrease of the iterate over each horizon and a bound on the closed-loop cost relative to the optimal infinite-horizon cost, which makes the trade-off between approximation accuracy, computational effort, and control performance explicit. Building on this decrease, we further characterize the inherent robustness of the controller against model errors: a condition on the effect of the model error over one horizon preserves asymptotic stability and a suboptimality bound. In the linear-quadratic case, implementing the optimal input of each horizon as a time-varying feedback yields input-to-state stability (ISS) \cite{sontag1989ISS,jiang2001ISSdiscrete} with respect to additive disturbances with explicit bounds, and recovers the classical robustness of the linear-quadratic regulator (LQR) as the errors vanish. Numerical studies on a linear-quadratic problem, where value iteration can be performed without approximation errors, and on a nonlinear system with known optimal value function, constructed via converse optimality, compare the predicted bounds with the actual closed-loop performance.

\textit{Notation:} $\norm{\cdot}$ denotes the Euclidean norm for vectors and the induced matrix norm for matrices. For a differentiable function $V$ and a vector field $H$, $L_HV(x)\coloneqq\nabla V(x)^\top H(x)$ is the Lie derivative of $V$ along $H$. The solution of \eqref{eq:system} at time $t$ under the input $u(\cdot)$ from the initial state $x$ is denoted by $x_u(t;x)$. A continuous, strictly increasing $\alpha:[0,\infty)\to[0,\infty)$ with $\alpha(0)=0$ is of class $\mathcal{K}$, and of class $\mathcal{K}_\infty$ if it is unbounded; a continuous $\beta:[0,\infty)^2\to[0,\infty)$ is of class $\mathcal{KL}$ if $\beta(\cdot,t)\in\mathcal{K}$ and $\beta(s,\cdot)$ is nonincreasing with limit $0$ for all $s,t\geq0$.

\section{Problem Setup and Preliminaries}
\label{sec:psetup-preliminaries}
We consider the continuous-time nonlinear system
\begin{align}
    \label{eq:system}
    \dot{x}(t) = f(x(t),u(t)), \quad t \in \R_{\geq 0},
\end{align}
where $x(t) \in \R^n$ denotes the state with initial
condition $x(0)=x_0$, and $u(t) \in \R^m$ is the control
input. The dynamics $f:\R^n \times \R^m \to \R^n$
are continuous and satisfy $f(0,0)=0$.

We consider a continuous stage cost of the form:
\begin{equation*}
    \ell(x,u) = Q(x) + u^\top R u,
    \quad
    Q:\R^n \to \R_{\geq 0},
    \quad R \succ 0.
\end{equation*}
The following assumption imposes a suitable lower bound on
the stage cost and is satisfied, for instance, by
quadratic stage costs.

\begin{asm}
    \label{asm:stage-cost}
    There exists $\underline{\alpha}_\ell \in \mathcal{K}_\infty$ such that $\underline{\alpha}_\ell(\|x\|)\leq \ell(x,0)$ for all $x \in \R^n$.
\end{asm}

We aim to determine the input $u$ that minimizes the functional
\begin{equation*}
    J_\infty \bigl(x,u(\cdot)\bigr)
    =
    \int_0^\infty
    \ell\bigl(x_u(\tau;x),u(\tau)\bigr)\,d\tau,
\end{equation*}
where we tacitly assume that such an input exists, and denote the optimal value by $V_\infty(x)\coloneqq\inf_{u(\cdot)}J_\infty\bigl(x,u(\cdot)\bigr)$. Since this problem is intractable for general nonlinear systems and costs, we approximate its solution by value iteration.

\subsection*{Value iteration in continuous time}
In discrete time, each iteration minimizes the stage cost plus the previous iterate at the successor state. Analogously, we replace one time step by a horizon $T>0$. Since performing such an iteration exactly is feasible only in particular cases such as the linear-quadratic regulator, we include approximation errors $e_j:\Omega\to\R$ and propose the following value iteration scheme, for all $x$ in a user-defined bounded domain $\Omega\subseteq\R^n$,
\begin{equation}
\label{eq:approx_ct_value_iteration}
\begin{aligned}
\hat{J}_j\bigl(x,u(\cdot)\bigr)
    &=
    \int_0^T
    \ell\bigl(x_u(\tau;x),u(\tau)\bigr)\,d\tau
    + \hat{V}_j\bigl(x_u(T;x)\bigr), \\
\hat{V}_{j+1}(x)
    &=
    \min_{u(\cdot)}
    \hat{J}_j\bigl(x,u(\cdot)\bigr) + e_j(x), \\
\hat{k}_j(\cdot;x)
    &=
    \operatorname*{arg\,min}_{u(\cdot)}
    \hat{J}_j\bigl(x,u(\cdot)\bigr).
\end{aligned}
\end{equation}
starting from $\hat{V}_0\equiv0$. For $e_j\equiv0$, we write $J_j$, $V_j$, and $k_j$ for the exact iteration. The terminal cost in \eqref{eq:approx_ct_value_iteration} is evaluated at the end points $x_u(T;x)$, which may leave $\Omega$. We therefore take each $\hat{V}_j$ to be defined on $\R^n$, as common approximators such as polynomials or neural networks are; \eqref{eq:approx_ct_value_iteration} prescribes its values only on $\Omega$. Under the conditions of Theorem~\ref{thm:h_stabilizing}, the analysis uses $\hat{V}_i$ only at end points of optimal trajectories, which lie in $\Omega$. We stop at an iteration $i$ at which consecutive iterates differ by at most a convergence tolerance $\delta:\Omega\to\R_{\geq0}$, i.e.,

\begin{equation}
    \label{eq:conv_tol_del_bound}
    | \hat{V}_{i+1}(x) - \hat{V}_i(x) | \leq \delta (x), \quad \forall x \in \Omega.
\end{equation}

\begin{asm}
    \label{asm:V1}
    There exists $\alpha_1\in\mathcal{K}$ such that $\alpha_1(\norm{x})\leq V_1(x)$ for all $x\in\Omega$.
\end{asm}
Since $V_1$ is the optimal cost over one horizon without terminal cost, Assumption~\ref{asm:V1} requires that no state can be steered to the origin within one horizon at arbitrarily small cost. It holds, for instance, under Assumption~\ref{asm:stage-cost} if $f$ satisfies the growth bound \eqref{eq:growth_bound} below: the state cannot leave a neighborhood of $x$ quickly without a costly input.

\section{Main Results}

\subsection{Asymptotic stability}

The following lemma relates the approximate iterates to the exact ones, cf.\ relaxed value iteration \cite[eq.~(12)]{lincoln2006relaxingDP}, and is only needed for the suboptimality bound relative to $V_\infty$.

\begin{lem}\label{lem:bounds}
Let $q\in\N$, $x\in\Omega$, and suppose there exists $c_e\geq0$ such that, for all $j<q$,
        \begin{equation}
        \label{eq:bounded_error_lemma}
            e_j(y) \leq c_e V_1(y), \quad \forall y\in \Omega.
        \end{equation}
If the optimal trajectory of the problem defining $V_q(x)$, i.e., over $[0,qT]$ without terminal cost, satisfies $x(nT)\in\Omega$ for $n=1,\dots,q-1$, then $\hat{V}_q(x)\leq(1+c_e)V_q(x)$.
\end{lem}

\begin{proof}
By dynamic programming, this trajectory on $[nT,(n+1)T]$ is generated by the exact minimizer $k_{q-n-1}$ from $y_n\coloneqq x(nT)$. We show $\hat V_{q-n}(y_n)\leq(1+c_e)V_{q-n}(y_n)$ backwards from $n=q$, where $\hat V_0=V_0$. Let $L$ be the running cost on $[nT,(n+1)T]$. Since $V_1$ is the minimal running cost over one horizon, $V_1(y_n)\leq L$, and \eqref{eq:bounded_error_lemma} gives
\begin{equation*}
	\begin{aligned}
	\hat{V}_{q-n}(y_n)&\leq L+\hat V_{q-n-1}(y_{n+1})+e_{q-n-1}(y_n)\\
	&\leq L+(1+c_e)V_{q-n-1}(y_{n+1})+c_eL\\
	&=(1+c_e)V_{q-n}(y_n).
	\end{aligned}
\end{equation*}
\end{proof}

\begin{thm}\label{thm:h_stabilizing}
		Consider the sampled-data controller $\kappa_i$ that, at the sampling instants $t_n\coloneqq nT$, $n\in\N\cup\{0\}$, solves the finite-horizon problem in \eqref{eq:approx_ct_value_iteration} with terminal cost $\hat{V}_i$ and applies its optimal input over the whole horizon, i.e.,
		\begin{equation}
		\label{eq:sampled_controller}
			u(t)=\hat{k}_i\bigl(t-t_n;x(t_n)\bigr), \quad t\in[t_n,t_{n+1}).
		\end{equation}
		Let $i\geq1$ and suppose there exist $c_e,c_{\delta}\geq0$ with $c_e+c_\delta<1$ such that, for all $x\in\Omega$,
        \begin{align}
            \label{eq:conv_tol_bound}
            &\delta(x) \leq c_\delta \int_0^T \ell \bigl(x_{\hat{k}_i}(\tau;x),\hat{k}_i(\tau;x)\bigr)\,d\tau, \\
            \label{eq:approx_error_bound}
            &|e_i(x)| \leq c_e V_1(x),
        \end{align}
		that $\hat V_i$ is bounded on $\Omega$ and continuous at the origin with $\hat V_i(0)=0$, that Assumption~\ref{asm:V1} holds, and that there exists $L_f>0$ such that
		\begin{equation}
		\label{eq:growth_bound}
			\norm{f(x,u)}\leq L_f\bigl(\norm{x}+\norm{u}\bigr)
		\end{equation}
		for all $u\in\R^m$ and all $x$ in a compact set $\mathcal{X}$ that contains the trajectories $x_{\hat{k}_i}(t;x)$, $t\in[0,T]$, from all $x\in\Omega$. Let $r>0$ be such that $x_{\hat k_i}(T;x)\in\Omega$ for all $x$ in the sublevel set $\mathcal{B}_r\coloneqq\{x\in\Omega:\hat V_i(x)\leq r\}$.
		Then $\kappa_i$ asymptotically stabilizes system \eqref{eq:system} for any initial state in $\mathcal{B}_r$: there exist $\beta\in\mathcal{KL}$ and $\gamma\in\mathcal{K}$ such that
		\begin{equation*}
		\norm{x(t)}\leq\gamma\bigl(\beta(\norm{x_0},\lfloor t/T\rfloor)\bigr), \quad t\geq0,
		\end{equation*}
        and the closed-loop cost satisfies $J_\infty(x_0,u(\cdot))\leq\hat V_i(x_0)/(1-c_e-c_\delta)$. If, moreover, the assumptions of Lemma~\ref{lem:bounds} hold for $q=i$ and $x=x_0$, then
        \begin{equation}
            \label{eq:suboptimality_bound}
            J_\infty\bigl(x_0,u(\cdot)\bigr) \leq \alpha_{e,\delta} V_\infty(x_0), \quad \alpha_{e,\delta}\coloneqq\frac{1+c_e}{1-c_e-c_\delta}.
        \end{equation}
	\end{thm}
The condition on the end points holds automatically if $\hat V_i$ is defined on $\R^n$ and its sublevel set $\{x\in\R^n:\hat V_i(x)\leq r\}$ lies in $\Omega$, since then $\hat V_i(x_{\hat k_i}(T;x))\leq\hat V_i(x)\leq r$ for $x\in\mathcal{B}_r$ by \eqref{eq:decay_rate} below.

	\begin{proof}
		\textit{Decrease.}
		For $x\in\Omega$, \eqref{eq:approx_ct_value_iteration} and \eqref{eq:conv_tol_del_bound} give
        \begin{equation}
    		\begin{aligned}
                \label{eq:decay_rate}
    			&\hat{V}_i\bigl(x_{\hat{k}_i}(T;x)\bigr) - \hat{V}_i(x) = \hat{V}_{i+1}(x)-\hat{V}_i(x) \\
                &\quad-\int_0^T
                \ell \bigl(x_{\hat{k}_i}(\tau;x),\hat{k}_i(\tau;x)\bigr)\,d\tau - e_i(x) \\
                &\overset{\eqref{eq:conv_tol_bound},\eqref{eq:approx_error_bound}}{\leq}
                (c_e+c_\delta-1)\int_0^T
                \ell \bigl(x_{\hat{k}_i}(\tau;x),\hat{k}_i(\tau;x)\bigr)\,d\tau,
    		\end{aligned}
        \end{equation}
		which is a relaxed dynamic programming inequality \cite{lincoln2006relaxingDP}. Here, $-e_i(x)\leq c_eV_1(x)$ and $V_1(x)$ is at most the running cost of $\hat{k}_i(\cdot;x)$. Under \eqref{eq:sampled_controller}, the closed loop on $[t_n,t_{n+1}]$ is the optimal trajectory from $x(t_n)$. For $x\in\mathcal{B}_r$, its end point lies in $\Omega$ by assumption, and $\hat V_i$ does not increase by \eqref{eq:decay_rate}, so the end point lies in $\mathcal{B}_r$. By induction, $x(t_n)\in\mathcal{B}_r$ for all $n$.

		\textit{Stability.}
		Let $c\coloneqq c_e+c_\delta$ and $L_n\coloneqq\int_{t_n}^{t_{n+1}}\ell\,d\tau$. Summing \eqref{eq:decay_rate} gives $(1-c)\sum_{n<N}L_n\leq\hat V_i(x_0)-\hat V_i(x(t_N))$. As $\hat V_i$ is bounded on $\Omega$, the sum converges, so $\alpha_1(\norm{x(t_n)})\leq V_1(x(t_n))\leq L_n\to0$. Thus $x(t_n)\to0$, $\hat V_i(x(t_N))\to0$, and
		\begin{equation}
		\label{eq:aposteriori}
		(1-c)\textstyle\sum_{n\geq0}L_n\leq\hat V_i(x_0).
		\end{equation}
		In particular, $\hat V_i(x)\geq(1-c)L_0\geq(1-c)\alpha_1(\norm{x})$ on $\mathcal{B}_r$. With $\varphi(s)\coloneqq\sup\{\hat V_i(x):x\in\mathcal{B}_r,\norm{x}\leq s\}$, $\overline\alpha(s)\coloneqq s+\tfrac1s\int_s^{2s}\varphi(r)\,dr$ is of class $\mathcal{K}$ and $\hat V_i(x)\leq\overline\alpha(\norm{x})$ on $\mathcal{B}_r$. Hence, by the discrete-time Lyapunov theorem \cite[Ch.~2]{grune2017NMPC}, $\norm{x(t_n)}\leq\beta(\norm{x_0},n)$ for some $\beta\in\mathcal{KL}$.

		\textit{Inter-sample bound.}
		Since $Q\geq0$, we have $\ell(x,u)\geq\lambda_{\min}(R)\norm{u}^2$. Moreover, $\hat V_i\geq0$ on $\mathcal{B}_r$ and \eqref{eq:decay_rate} give $L_n\leq\hat V_i(x(t_n))/(1-c)$. With $s_n\coloneqq\norm{x(t_n)}$ and $\eta(s)\coloneqq\overline\alpha(s)/((1-c)\lambda_{\min}(R))$, the input energy over one horizon thus satisfies
		\begin{equation*}
			\int_{t_n}^{t_{n+1}}\norm{u(\tau)}^2d\tau\leq\frac{L_n}{\lambda_{\min}(R)}\leq\eta(s_n).
		\end{equation*}
		For $t\in[t_n,t_{n+1})$, the closed-loop trajectory lies in $\mathcal{X}$, so \eqref{eq:growth_bound} and the Cauchy--Schwarz inequality give
		\begin{equation*}
			\begin{aligned}
			\norm{x(t)}&\leq s_n+L_f\int_{t_n}^{t}\bigl(\norm{x(\tau)}+\norm{u(\tau)}\bigr)\,d\tau\\
			&\leq s_n+L_f\sqrt{T\eta(s_n)}+L_f\int_{t_n}^{t}\norm{x(\tau)}\,d\tau.
			\end{aligned}
		\end{equation*}
		Gr\"onwall's inequality then yields
		\begin{equation*}
			\norm{x(t)}\leq e^{L_fT}\Bigl(s_n+L_f\sqrt{T\eta(s_n)}\Bigr)\eqqcolon\gamma(s_n),
		\end{equation*}
		where $\gamma\in\mathcal{K}$ since $\overline\alpha\in\mathcal{K}$. Thus, $\norm{x(t)}\leq\gamma(\beta(\norm{x_0},\lfloor t/T\rfloor))$, i.e., the continuous-time closed loop of this sampled-data scheme \cite{fontes2001sampledMPC} is asymptotically stable.

		\textit{Suboptimality.}
		The closed-loop cost is $\sum_{n\geq0}L_n$, so \eqref{eq:aposteriori} is the first bound. Lemma~\ref{lem:bounds} and $V_i\leq V_\infty$ give $\hat V_i(x_0)\leq(1+c_e)V_\infty(x_0)$, hence \eqref{eq:suboptimality_bound}.
	\end{proof}

\begin{rem}\label{rem:sampling}
Conditions \eqref{eq:conv_tol_bound}, \eqref{eq:approx_error_bound}, and the end-point condition involve only the final iterates, but must hold on $\Omega$ and $\mathcal{B}_r$, and evaluating them at a state requires solving a finite-horizon optimal control problem. For general nonlinear systems, they can thus only be estimated from finitely many states, as in Section~\ref{sec:nonlinear}.
\end{rem}

\subsection{Inherent robustness}
We investigate the inherent robustness of the controller $\kappa_i$ with respect to model errors. Let the true system be
\begin{equation}
\label{eq:true_system}
    \dot{x}(t)=f_a\bigl(x(t),u(t)\bigr),
\end{equation}
while the controller still solves the finite-horizon problem with the nominal model $f$. For a sampling instant $t_n$ with state $x_n\coloneqq x(t_n)$, let $x_p$ denote the \textit{predicted} trajectory, i.e., $x_p(t)\coloneqq x_{\hat{k}_i}(t-t_n;x_n)$, and $x_a$ the \textit{actual} trajectory of \eqref{eq:true_system} on $[t_n,t_{n+1}]$, both starting from $x_n$ and driven by the input of the controller. We define the trajectory error $\Delta(t)\coloneqq x_a(t)-x_p(t)$, with $\Delta(t_n)=0$, and the predicted and actual costs over one horizon
\begin{equation*}
\begin{aligned}
    L_p(x_n)&\coloneqq\int_{t_n}^{t_{n+1}} \ell\bigl(x_p(\tau),u(\tau)\bigr)\,d\tau,\\
    L_a(x_n)&\coloneqq\int_{t_n}^{t_{n+1}} \ell\bigl(x_a(\tau),u(\tau)\bigr)\,d\tau .
\end{aligned}
\end{equation*}

\begin{lem}\label{lem:robust_stability}
    Suppose the assumptions of Theorem~\ref{thm:h_stabilizing} hold, Assumption~\ref{asm:V1} and \eqref{eq:growth_bound} also hold for $f_a$, the latter on a compact set that contains the actual trajectories, and let $c\coloneqq c_e+c_\delta<1$. If there exists $c_\Delta\in[0,1-c)$ such that, for all $x_n\in\mathcal{B}_r$, with $\mathcal{B}_r$ as in Theorem~\ref{thm:h_stabilizing}, the actual end point $x_a(t_{n+1})$ lies in $\Omega$ and
    \begin{equation}
    \label{eq:inherent_robustness_cond}
        \begin{aligned}
        &\hat{V}_i\bigl(x_a(t_{n+1})\bigr)-\hat{V}_i\bigl(x_p(t_{n+1})\bigr)\\
        &\quad+(1-c)\bigl(L_a(x_n)-L_p(x_n)\bigr)\leq c_\Delta L_a(x_n),
        \end{aligned}
    \end{equation}
    then $\kappa_i$ asymptotically stabilizes the true system \eqref{eq:true_system} for any initial state in $\mathcal{B}_r$, and
    \begin{equation}
    \label{eq:robust_decrease}
        \begin{aligned}
        \hat{V}_i\bigl(x(t_{n+1})\bigr)-\hat{V}_i\bigl(x(t_n)\bigr)
        \leq-(1-c-c_\Delta)\,L_a\bigl(x(t_n)\bigr)
        \end{aligned}
    \end{equation}
    for all $n$. If, moreover, the assumptions of Lemma~\ref{lem:bounds} hold for $q=i$ and $x=x_0$, the closed-loop cost of the true system satisfies
    \begin{equation*}
        J_\infty^a(x_0)\leq\frac{1+c_e}{1-c_e-c_\delta-c_\Delta}\,V_\infty(x_0),
    \end{equation*}
    with $V_\infty$ the nominal optimal cost.
\end{lem}
\begin{proof}
    For $x_n\in\mathcal{B}_r$, let $x_a^+\coloneqq x_a(t_{n+1})$ and $x_p^+\coloneqq x_p(t_{n+1})$. Splitting the actual change of $\hat{V}_i$ into the predicted one and a correction, and using the decrease \eqref{eq:decay_rate} along the predicted trajectory,
    \begin{equation*}
    \begin{aligned}
        &\hat{V}_i(x_a^+)-\hat{V}_i(x_n)\\
        &=\hat{V}_i(x_p^+)-\hat{V}_i(x_n)+\hat{V}_i(x_a^+)-\hat{V}_i(x_p^+)\\
        &\overset{\eqref{eq:decay_rate}}{\leq}-(1-c)L_p(x_n)+\hat{V}_i(x_a^+)-\hat{V}_i(x_p^+)\\
        &=-(1-c)L_a(x_n)+\hat{V}_i(x_a^+)-\hat{V}_i(x_p^+)\\
        &\quad+(1-c)\bigl(L_a(x_n)-L_p(x_n)\bigr)\\
        &\overset{\eqref{eq:inherent_robustness_cond}}{\leq}-(1-c-c_\Delta)L_a(x_n),
    \end{aligned}
    \end{equation*}
    which is \eqref{eq:robust_decrease}, since $x(t_{n+1})=x_a^+$. As $c_\Delta<1-c$ and $x_a(t_{n+1})\in\Omega$, $\mathcal{B}_r$ is invariant at the sampling instants. The actual cost over one horizon is bounded below by the optimal cost over one horizon of the true system without terminal cost, which admits a lower bound in $\mathcal{K}$ by Assumption~\ref{asm:V1} for $f_a$. The bounds on $\hat{V}_i$ and the input energy are unchanged, and the inter-sample bound follows as before with $f_a$ instead of $f$. Asymptotic stability then follows as in the proof of Theorem~\ref{thm:h_stabilizing}. Summing \eqref{eq:robust_decrease} over $n$ and using $\hat{V}_i\geq0$ and, as in Theorem~\ref{thm:h_stabilizing}, $\hat V_i(x_0)\leq(1+c_e)V_\infty(x_0)$ gives the bound on $J_\infty^a$.
\end{proof}

\subsection{ISS in the linear-quadratic case}\label{sec:lqr}
In the linear-quadratic case, the system dynamics and the stage cost have the form
\begin{equation*}
    \ell(x,u) = x^\top Q x + u^\top R u, \quad f(x,u) = Ax + Bu,
\end{equation*}
with $Q\succ0$ and $R\succ0$, and the iterates are $\hat{V}_j(x)=x^\top P_jx$. Over one horizon, the finite-horizon problem with terminal cost $\hat{V}_j$ is solved by the Riccati equation $-\dot{P}=A^\top P+PA-PBR^{-1}B^\top P+Q$ on $[0,T]$ with $P(T)=P_j$; its optimal value $x^\top P(0)x$ differs from $\hat{V}_{j+1}(x)$ by $e_j(x)$. Instead of applying the optimal input as an open-loop signal, we implement it as the time-varying feedback
\begin{equation}
\label{eq:lqr_window_feedback}
    u(t)=-K(t-t_n)\,x(t),\quad t\in[t_n,t_{n+1}),
\end{equation}
with $K(s)\coloneqq R^{-1}B^\top P(s)$. For the nominal model, \eqref{eq:lqr_window_feedback} generates exactly the optimal trajectory, so Theorem~\ref{thm:h_stabilizing} applies unchanged. For the controller $\kappa_i$, i.e., $P(T)=P_i$, let $\Phi(t,s)$ be the transition matrix of $A-BK(t)$, $\Phi_i\coloneqq\Phi(T,0)$, and $S_i\coloneqq P(0)-\Phi_i^\top P_i\Phi_i\succ0$, so that $L_p(x)=x^\top S_ix$.

\begin{crl}\label{crl:lqr_iss}
    Suppose \eqref{eq:conv_tol_bound} and \eqref{eq:approx_error_bound} hold with $\Omega=\R^n$ and $c_e+c_\delta<1$. Let $f_a(x,u)=Ax+Bu+d(t)$ with $\norm{d(t)}\leq\overline{d}$ for all $t\geq0$, and let $\kappa_i$ be implemented by \eqref{eq:lqr_window_feedback}. Then, with $\gamma_\Delta\coloneqq\int_0^T\norm{\Phi(T,s)}\,ds$, the states $x_n\coloneqq x(t_n)$ of the closed loop satisfy, for all $n\geq0$,
    \begin{equation*}
    \begin{aligned}
        &\hat{V}_i(x_{n+1})-\hat{V}_i(x_n)\leq-\tfrac{1}{2}(1-c_e-c_\delta)\lambda_{\min}(S_i)\norm{x_n}^2\\
        &\quad+\Bigl(\tfrac{2\norm{\Phi_i^\top P_i}^2}{(1-c_e-c_\delta)\lambda_{\min}(S_i)}+\norm{P_i}\Bigr)\gamma_\Delta^2\,\overline{d}^2,
    \end{aligned}
    \end{equation*}
    and the closed loop is ISS \cite{sontag1989ISS} with respect to $d$ for all initial states.
\end{crl}
\begin{proof}
    Under \eqref{eq:lqr_window_feedback}, $\dot{\Delta}=(A-BK(t-t_n))\Delta+d$, so $x_{n+1}=\Phi_ix_n+\Delta_{n+1}$ with $\Delta_{n+1}\coloneqq\Delta(t_{n+1})$ and $\norm{\Delta_{n+1}}\leq\gamma_\Delta\overline{d}$. Hence,
    \begin{equation*}
    \begin{aligned}
        &\hat{V}_i(x_{n+1})-\hat{V}_i(x_n)=x_n^\top\bigl(\Phi_i^\top P_i\Phi_i-P_i\bigr)x_n\\
        &\qquad+2x_n^\top\Phi_i^\top P_i\Delta_{n+1}+\Delta_{n+1}^\top P_i\Delta_{n+1}\\
        &\overset{\eqref{eq:decay_rate}}{\leq}-(1-c_e-c_\delta)\,x_n^\top S_ix_n\\
        &\qquad+2\norm{\Phi_i^\top P_i}\norm{x_n}\gamma_\Delta\overline{d}+\norm{P_i}\gamma_\Delta^2\overline{d}^2,
    \end{aligned}
    \end{equation*}
    where, since $\Phi_ix_n$ is the predicted end point and $L_p(x_n)=x_n^\top S_ix_n$, \eqref{eq:decay_rate} bounds the first term. Using $x_n^\top S_ix_n\geq\lambda_{\min}(S_i)\norm{x_n}^2$ and Young's inequality $2ab\leq\varepsilon a^2+b^2/\varepsilon$ with $\varepsilon=\tfrac12(1-c_e-c_\delta)\lambda_{\min}(S_i)$ gives the stated inequality, which makes the quadratic function $\hat{V}_i$ a global ISS-Lyapunov function for the sampled closed loop \cite{jiang2001ISSdiscrete}. Between sampling instants, $x(t)=\Phi(t-t_n,0)x_n+\Delta(t)$ gives $\norm{x(t)}\leq M_\Phi(\norm{x_n}+T\overline{d})$ with $M_\Phi\coloneqq\max_{0\leq\tau\leq s\leq T}\norm{\Phi(s,\tau)}$, which extends ISS to all $t\geq0$.
\end{proof}

Without approximation errors, $c_\delta\to0$ and $P_i$ converges, as $i\to\infty$, to a fixed point of the Riccati map over one horizon, which, for $(A,B)$ stabilizable and $\hat{V}_0\equiv0$, is the stabilizing solution $P_\infty$ of the algebraic Riccati equation. The Riccati solution with terminal value $P_\infty$ is constant, so $K(t)\equiv R^{-1}B^\top P_\infty$ and \eqref{eq:lqr_window_feedback} reduces to the infinite-horizon LQR, whose classical ISS property with respect to additive disturbances is thus recovered in the limit.

\section{Numerical Studies}

\subsection{The linear-quadratic case}
We consider the double integrator with the quadratic cost
\begin{equation*}
A=\begin{bmatrix}0&1\\0&0\end{bmatrix},\quad
B=\begin{bmatrix}0\\1\end{bmatrix},\quad
Q=I_2,\quad R=1,
\end{equation*}
and set $T=1$. Each iteration integrates the Riccati equation of Section~\ref{sec:lqr} backward from $P(T)=P_j$ and sets $P_{j+1}=P(0)$, i.e., we solve the finite-horizon optimal control problem (OCP) exactly. We evaluate the controller $\kappa_j$ of Theorem~\ref{thm:h_stabilizing} with terminal cost $\widehat V_j$, implemented by \eqref{eq:lqr_window_feedback}. As all quantities are quadratic and $P_j\succ0$ for $j\geq1$, the constants below hold on every ball $\Omega$, so Theorem~\ref{thm:h_stabilizing} applies to every initial state by choosing $\Omega$ large.

With $\Phi_j$ and $S_j$ as in Section~\ref{sec:lqr}, the OCP solution from $x$ ends at $\Phi_jx$ with running cost $x^\top S_jx$, so
\[
 c_\delta=\sup_{x\neq0}\frac{|x^\top(P_{j+1}-P_j)x|}{x^\top S_jx}
 =\rho\bigl(S_j^{-1}(P_{j+1}-P_j)\bigr),
\]
where $\rho$ denotes the spectral radius, since $S_j\succ0$ turns the ratio into a generalized Rayleigh quotient.

Regarding suboptimality, under $\kappa_j$ the state at the sampling instants evolves as $x(t_{n+1})=\Phi_jx(t_n)$, and each horizon costs $x(t_n)^\top S_jx(t_n)$. Hence, the closed-loop cost from $x_0$ is $J^{\mathrm{cl}}_j(x_0)=x_0^\top X_jx_0$, where $X_j$ solves the discrete Lyapunov equation $X_j=S_j+\Phi_j^\top X_j\Phi_j$. Since $V_\infty(x_0)=x_0^\top P_\infty x_0$ is quadratic as well, we have
\[
 \sup_{x_0\neq0}\frac{J^{\mathrm{cl}}_j(x_0)}{V_\infty(x_0)}
 =\lambda_{\max}\bigl(P_\infty^{-1}X_j\bigr).
\]

\begin{figure}[t]
\centering
\begin{tikzpicture}
\begin{axis}[
  width=0.97\columnwidth, height=0.45\columnwidth,
  xmin=0, xmax=19,
  ymode=log, ymin=1e-14, ymax=10, ytick={1e-12,1e-8,1e-4,1},
  xlabel={Value iteration $j$}, ylabel={Value of the constant},
  tick label style={font=\footnotesize}, label style={font=\small},
  grid=major, grid style={gray!20},
  every axis plot/.append style={thick},
  legend pos=south west, legend style={font=\footnotesize, draw=none, fill=none},
]
\addplot[black] table[x expr=\thisrow{k}-1, y=c_delta] {data2plot/lqr_constants.dat};
\addlegendentry{$c_\delta$, LQR}
\addplot[black, dashed] table[x expr=\thisrow{k}-1, y=c_delta] {data2plot/nonlinear_constants.dat};
\addlegendentry{$c_\delta$, nonlinear}
\addplot[black, dotted] table[x expr=\thisrow{k}-1, y=c_e_bar] {data2plot/nonlinear_constants.dat};
\addlegendentry{$c_e$, nonlinear}
\end{axis}
\end{tikzpicture}
\caption{Constants over the value iterations (logarithmic scale). For LQR, $c_e$ is at rounding level.}
\label{fig:constants_lqr}\label{fig:constants_nonlinear}
\end{figure}
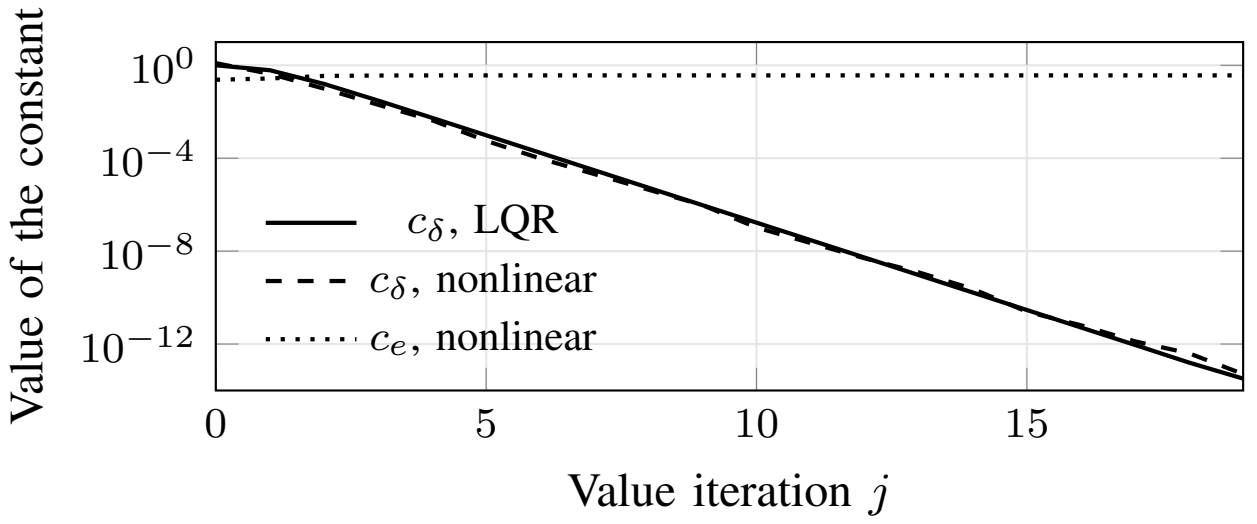

As shown in Fig.~\ref{fig:constants_lqr}, the constant $c_\delta$ decreases geometrically from $j=3$, by a factor of about $0.18$ per iteration, and is at rounding level from $j=20$ (cropped).
Accordingly, $\kappa_0$ has no guarantee, since $c_\delta+c_e\geq1$. This holds for every iteration started from zero: as $\widehat V_0\equiv0$, \eqref{eq:approx_error_bound} gives $\delta\geq\widehat V_1\geq(1-c_e)L$ with $L$ the running cost of $\hat k_0$, so $c_\delta\geq1-c_e$.

We measure the error by the worst-case
relative deviation
$\varepsilon_j=\sup_{x\neq0}|\widehat V_j(x)-V_\infty(x)|/V_\infty(x)$,
so that $|\widehat V_j(x)-V_\infty(x)|\le\varepsilon_jV_\infty(x)$ for every
state. By the same argument as for $c_\delta$, we can calculate it exactly as
$\varepsilon_j=\rho(P_\infty^{-1}P_j-I)$. Since $P_0=0$ and the Riccati map over one horizon is monotone, the iterates
increase and approach $P_\infty$ from below.

The exact worst-case closed-loop cost never exceeds the predicted factor. The bound is conservative for early
controllers: at $j=1$ it predicts a ratio of at most $2.59$, while the
closed-loop cost exceeds $V_\infty$ by at most $6.8\%$. The controller $\kappa_0$, which has no guarantee, has a worst-case ratio of $4.55$: with zero terminal cost, the optimal input of each horizon ignores the state at its end. The iterates converge to $P_\infty$.

\begin{table}[t]
\centering
\caption{Constants and suboptimality results for the LQR case}
\label{tab:lqr}
\small
\begin{tabular}{rcccc}
\hline
$j$ & $c_\delta$ & $\varepsilon_j$ & $\alpha_{e,\delta}$ &
$\sup_{x_0}J^{\mathrm{cl}}_j/V_\infty$ \\
\hline
0  & 1      & 1    & --     & 4.5481 \\
1  & 0.61   & 0.50 & 2.5867 & 1.0681 \\
2  & 0.16   & 0.13 & 1.1879 & 1.0022 \\
3  & 0.030  & 0.026 & 1.0314 & 1.0001 \\
4  & 0.0055 & 0.0046 & 1.0055 & 1.0000 \\
49 & $\approx0$ & $\approx0$ & 1.0000 & 1.0000 \\
\hline
\end{tabular}
\end{table}

\subsection{Nonlinear system benchmark using converse optimality}
\label{sec:nonlinear}

For general nonlinear systems, the optimal value function is not known in closed form. To obtain a nonlinear test problem with known optimal value function $V_\infty$, we use the nonlinear benchmark generation of \cite{yamerenko2025benchmark}: for a prescribed $V_\infty$, a vector field $H$, and an input field $G$, the system $\dot{x}=\tfrac14G(x)\bigl(L_GV_\infty(x)\bigr)^\top-H(x)+G(x)u$ with stage cost $\ell(x,u)=L_HV_\infty(x)+u^2$ has the optimal value function $V_\infty$. We choose
\begin{equation*}
    V_\infty(x)=\tfrac{1}{2}x_1^2+\tfrac{1}{4}x_2^4+\tfrac{1}{2}(1-\cos x_1)+\tfrac{1}{2}\log\cosh x_2,
\end{equation*}
which combines polynomial and non-polynomial nonlinearities, $H(x)=(I_2+\Psi)\nabla V_\infty(x)$ with the skew-symmetric $\Psi=\left[\begin{smallmatrix}0&-1\\1&0\end{smallmatrix}\right]$, so that $L_HV_\infty=\norm{\nabla V_\infty}^2$, and $G(x)=\bigl[1+\tfrac12\sin x_2,\;1+x_1^2\bigr]^\top$. The stage cost is then
\begin{equation*}
    \ell(x,u)=\bigl(x_1+\tfrac{1}{2}\sin x_1\bigr)^2+\bigl(x_2^3+\tfrac{1}{2}\tanh x_2\bigr)^2+u^2.
\end{equation*}

Each OCP ($T=0.5$) is solved by Legendre collocation of degree two with $50$ intervals of length
$h=0.01$ and piecewise-constant inputs, using IPOPT via CasADi initialized with the zero input. Starting from $\widehat V_0=0$, $\widehat V_{j+1}$ is fitted by least
squares to the optimal OCP values, with terminal cost $\widehat V_j$,
using all monomials of degrees two to four. We choose $200$
fitting states in the box $|x_1|\le\pi$, $|x_2|\le2$. Of those, $60\%$ are
uniform in the box, $20\%$ on its boundary, and $20\%$ along random
directions at $10^{-3}$ to $10^{-1}$ of the distance to the boundary,
log-uniformly.

To determine $c_\delta$ and $c_e$, we evaluate $|\widehat V_{j+1}(x)-\widehat V_j(x)|$ relative to the running cost of the OCP solution from $x$, and $|e_j(x)|$ relative to $V_1(x)$, the OCP value of the first iteration, over an independent validation set of $200$ states drawn like the
fitting states and determine the worst case. As Lemma~\ref{lem:bounds} involves $e_0,\dots,e_{j-1}$, $c_e$ covers all errors $e_0,\dots,e_j$; Theorem~\ref{thm:h_stabilizing} alone needs only $e_j$.

To determine the worst-case closed-loop cost, we simulate the sampled-data controller $\kappa_j$ of Theorem~\ref{thm:h_stabilizing} with terminal cost $\widehat V_j$: every $T$, it solves the OCP and applies the piecewise-constant optimal input over the whole horizon, for $5\,\mathrm{s}$. With $\Omega$ the box, we start from 200 states in $\mathcal{B}_{r_j}$, with $r_j$ the minimum of $\widehat V_j$ on the box boundary. At all validation states, the end points of the computed OCP solutions remained in the box, as required by Theorem~\ref{thm:h_stabilizing}, and so did, for the iterations in Table~\ref{tab:nonlinear}, those of the exact minimizers required by Lemma~\ref{lem:bounds}, i.e., by dynamic programming, the states at $T,\dots,(j-1)T$ of the OCP over $[0,jT]$ without terminal cost. The starts lie on uniformly random rays from the origin, shared by all controllers, at a fraction $s$ of the distance to the boundary of the sublevel set: $s=\sqrt{U}$ with $U$ uniform on $(0,1]$ for $60\%$ of them, $s=1$ for $20\%$, and $s$ log-uniform in $[10^{-3},10^{-1}]$ for the remaining $20\%$. The measured ratio is the closed-loop cost divided by $V_\infty(x_0)$ and reported in Table~\ref{tab:nonlinear}; $\kappa_0$, which has no guarantee, is not simulated. In all runs, \eqref{eq:decay_rate} held at every sampling instant, and the cost after $5\,\mathrm{s}$, bounded by $\widehat V_j(x(5\,\mathrm{s}))/(1-c_\delta-c_e)$ via \eqref{eq:aposteriori}, is below $2\cdot10^{-5}$ of the simulated cost.

As shown in Fig.~\ref{fig:constants_nonlinear}, the constant $c_\delta$ decreases geometrically and is negligible compared with $c_e$ from about $j=9$. Hence, after a few iterations,
$c_\delta+c_e$ is dominated by the approximation error. It is below one from $j=1$ on.

The constant $c_e$ reaches
$0.368$ at $j=5$ and stays there, which limits
$1-c_\delta-c_e$ to $0.63$ and the predicted factor $\alpha_{e,\delta}$ to $2.17$. Regarding the approximation error with respect to the optimal value function, we
calculate the worst-case relative deviation
$\varepsilon_j=\max_{x\neq0}|\widehat V_j(x)-V_\infty(x)|/V_\infty(x)$ over a
$41\times41$ grid on the box.  This worst case is attained at few states:
for example, for $\widehat V_{49}$, the median relative deviation is $3\%$, and
$\widehat V_{49}/V_\infty$ ranges from $0.68$ to $1.09$, so the fitted
value also overestimates $V_\infty$ at some states.

The measured suboptimality ratios in Table~\ref{tab:nonlinear} never exceed the predicted factor, but the bound is conservative: they stay below $1.020$, while $\alpha_{e,\delta}\geq2.17$.

\begin{table}[t]
\centering
\caption{Constants and suboptimality results for the nonlinear case}
\label{tab:nonlinear}
\small
\setlength{\tabcolsep}{4pt}
\begin{tabular}{rcccccc}
\hline
$j$ & $c_\delta$ & $c_e$ & $\varepsilon_j$ & $1-c_\delta-c_e$ & $\alpha_{e,\delta}$ &
$\max J^{\mathrm{cl}}_j/V_\infty$ \\
\hline
0  & 1.24   & 0.239 & 1    & $-0.477$ & --   & -- \\
1  & 0.43   & 0.271 & 0.53 & 0.300    & 4.23 & 1.0199 \\
2  & 0.099  & 0.349 & 0.36 & 0.552    & 2.45 & 1.0086 \\
3  & 0.020  & 0.363 & 0.33 & 0.617    & 2.21 & 1.0068 \\
4  & 0.0043 & 0.367 & 0.32 & 0.628    & 2.18 & 1.0064 \\
49 & $\approx0$ & 0.368 & 0.32 & 0.632 & 2.17 & 1.0063 \\
\hline
\end{tabular}
\end{table}

\section{Conclusion}
The results show that closed-loop guarantees for continuous-time value iteration do not require convergence of the iterates: a single iteration step whose convergence tolerance and approximation error are small relative to the cost over one horizon suffices for stability, a suboptimality bound, and inherent robustness of the associated sampled-data controller. Measuring both relative to the cost over a horizon, rather than over a time step, keeps these conditions meaningful in continuous time. In the numerical studies, the suboptimality bounds were never violated but proved conservative. In the linear-quadratic case, the conditions can be checked exactly for all states. For general nonlinear systems, they can only be evaluated at sampled states, as in the nonlinear example. This is the practical setting we target: deriving guarantees from such samples is the main direction of future work, together with using the approximate value function as a terminal cost in model predictive control to guarantee stability, constraint satisfaction, and performance requirements of the closed loop.

\bibliographystyle{IEEEtran}
\bibliography{bibliography}

\end{document}